\documentclass[sigconf,natbib=true]{acmart}

\AtBeginDocument{%
  \providecommand\BibTeX{{%
    \normalfont B\kern-0.5em{\scshape i\kern-0.25em b}\kern-0.8em\TeX}}}

\copyrightyear{2023} 
\acmYear{2023} 
\setcopyright{acmlicensed}\acmConference[RecSys '23]{Seventeenth ACM Conference on Recommender Systems}{September 18--22, 2023}{Singapore, Singapore}
\acmBooktitle{Seventeenth ACM Conference on Recommender Systems (RecSys '23), September 18--22, 2023, Singapore, Singapore}
\acmPrice{15.00}
\acmDOI{10.1145/3604915.3608831}
\acmISBN{979-8-4007-0241-9/23/09}

\usepackage[linesnumbered,ruled]{algorithm2e}
\usepackage{amsmath}
\usepackage{subfigure}
\usepackage{bm}
\usepackage{mathrsfs}
\usepackage{color}
\usepackage{multirow}
\usepackage{graphicx}

\usepackage{xspace}
\usepackage{caption} 
\usepackage{array}
\usepackage{subfigure}
\usepackage{svg}
\usepackage{cleveref}
\usepackage{booktabs}
\usepackage{enumitem}
\usepackage[T1]{fontenc}

\begin{document}

\title{Enhancing Transformers without Self-supervised Learning: A Loss Landscape Perspective in Sequential Recommendation
}


\author{Vivian Lai}
\email{viv.lai@visa.com}
\affiliation{%
  \institution{Visa Research}
  \city{Palo Alto}
  \country{USA}}

\author{Huiyuan Chen}
\email{hchen@visa.com}
\affiliation{%
  \institution{Visa Research}
  \city{Palo Alto}
  \country{USA}}

\author{Chin-Chia Michael Yeh}
\email{miyeh@visa.com}
\affiliation{%
  \institution{Visa Research}
  \city{Palo Alto}
  \country{USA}}

\author{Minghua Xu}
\email{mixu@visa.com}
\affiliation{%
  \institution{Visa Research}
  \city{Palo Alto}
  \country{USA}}

\author{Yiwei Cai}
\email{yicai@visa.com}
\affiliation{%
  \institution{Visa Research}
  \city{Palo Alto}
  \country{USA}}

\author{Hao Yang}
\email{haoyang@visa.com}
\affiliation{%
  \institution{Visa Research}
  \city{Palo Alto}
  \country{USA}}

\renewcommand{\shortauthors}{ViV Lai et al.}

\renewcommand{\shorttitle}{Enhancing Transformer without Self-supervised Learning}

\newcommand{\viv}[1]{\textcolor{red}{[#1 ---\textsc{viv}]}}
\newcommand{\hy}[1]{\textcolor{blue}{[#1 ---\textsc{huiyuan}]}}
\newcommand{\para}[1]{\noindent{\bf #1}}

\begin{abstract}
Transformer and its variants are a powerful class of architectures for sequential recommendation, owing to their ability of capturing a user's dynamic interests from their past interactions. Despite their success, Transformer-based models often require the optimization of a large number of parameters, making them difficult to train from sparse data in sequential recommendation. To address the problem of data sparsity, previous studies have utilized self-supervised learning to enhance Transformers, such as pre-training embeddings from item attributes or contrastive data augmentations. However, these approaches encounter several training issues, including initialization sensitivity, manual data augmentations, and large batch-size memory bottlenecks.

In this work, we investigate Transformers from the perspective of loss geometry, aiming to enhance the models' data efficiency and generalization in sequential recommendation. We observe that Transformers (e.g., SASRec) can converge to extremely sharp local minima if not adequately regularized. Inspired by the recent Sharpness-Aware Minimization (SAM), we propose SAMRec, which significantly improves the accuracy and robustness of sequential recommendation. SAMRec performs comparably to state-of-the-art self-supervised Transformers, such as S$^3$Rec and CL4SRec, without the need for pre-training or strong data augmentations. 
\end{abstract}


\maketitle

\section{Introduction}
\label{sec:intro}

Recommender systems play a significant role in online shopping platforms to enhance user satisfaction~\cite{koren2009matrix,chen2021tops}. Various methods have been proposed to personalize user browsing behaviors, making recommendations more effective~\cite{rendle2009bpr,HidasiKBT15,yeh2022embedding,lai2023selective,lai2022human}. However, the interests of users are inherently dynamic and evolve over time. This presents a new challenge for platforms to provide appropriate recommendations.  Sequential recommendation is one of the most commonly used methods, which aims to model the temporal dependencies of users' historical interactions and capture their dynamic preferences~\cite{kang2018self,sun2019bert4rec,li2020time}. Several successful sequential recommenders include Markov Chains \citep{rendle2010factorizing}, Recurrent Neural Networks \citep{HidasiKBT15}, and Convolutional Neural Networks \citep{tang2018personalized}.

In recent years, the Transformer architecture has demonstrated impressive capabilities in various tasks, including machine translation and computer vision \cite{vaswani2017attention, devlin2019bert, chen2022when}. The core idea behind Transformers is the self-attention mechanism, which models pairwise dependencies without regard to their distance in item sequences, enabling more parallelized training. Inspired by the success of Transformers, several self-attentive sequential recommenders have been proposed, achieving state-of-the-art performance in sequential recommendation \cite{kang2018self, sun2019bert4rec, li2020time, chen2022denoising, de2021transformers4rec}. For instance, SASRec \citep{kang2018self} takes advantage of this architecture, and its self-attention mechanism assigns weights to each previous item at every step, exhibiting impressive performance in several benchmark datasets. Bert4Rec \cite{sun2019bert4rec} further extends SASRec as a bidirectional self-attention encoder to exploit any pairwise item dependencies within the sequences. Recently, Transformers4Rec~\cite{de2021transformers4rec} has been developed to connect the fields of NLP and RecSys by integrating with one the most popular NLP frameworks, HuggingFace Transformers\footnote{https://github.com/huggingface/transformers}.

Despite their promising results, current Transformers models still face challenges with sparsity and often result in poor generalization in sequential recommendation~\cite{xie2022contrastive,zhou2020s3,ma2020disentangled}. Most Transformer-based models depend on observed user-item interactions to capture item-item dependencies, but the sparsity of these interactions makes it insufficient to learn sequential dependencies~\cite{liu2021augmenting}. Additionally, these models often result in poor generalization ability, especially for out-of-distribution items due to uncertain distribution shifts between the training and test data~\cite{chen2022adversariall,chen2022graphh,wang2022improving}. 

To address the aforementioned issues, one potential solution is the self-supervised learning, which allows the models to learn from the intrinsic structure of raw data~\cite{zhou2020s3,ma2020disentangled,qiu2022contrastive,xie2022contrastive,chen2022intent,wang2023federated}. For example, S$^3$Rec~\cite{zhou2020s3} uses four self-supervised objectives to study multiple correlations, including item-attribute, sequence-item, sequence-attribute, and sequence-segment relationships. CL4SRec~\cite{xie2022contrastive} improves the user representation by maximizing the agreement between different views of the same user sequence. However, these approaches may encounter training issues, such as initialization sensitivity, the requirement of additional item features, manual data augmentation (e.g., item crop, item mask, and item reorder), and large batch-size memory bottlenecks. A natural follow-up question would be: are there any alternative strategies to improve model generalization without relying on self-supervised signals?

Fortunately,  recent studies have shown a strong correlation between the sharpness of the training loss and the generalization error~\cite{foret2021sharpnessaware, Jiang2020Fantastic, keskar2017on, li2018visualizing}. Inspired by this finding, we investigate the loss landscapes of Transformer-based   models, such as SASRec, S$^3$Rec, and CL4SRec, in order to gain more insights into their training dynamics. By visualizing their loss landscapes (see Figure~\ref{fig1}), we reveal that   SASRec is likely to converge to extremely sharp local minima, while S$^3$Rec and CL4SRec converge to smoother minima as a result of utilizing self-supervised learning approach.

Motivated by recent developments in Sharpness-Aware Minimization (SAM)~\cite{foret2021sharpnessaware, kwon2021asam}, we propose a new sequential recommender framework, called SAMRec, which is designed to smooth the loss geometry of Transformers during training explicitly. To achieve this, SAMRec directly penalizes sharp minima and biases the convergence towards flatter regions by using a mini-max optimization approach: the minimization problem is used to perform standard model training, while the maximization problem helps the model escape sharp minima. We extensively evaluate SAMRec on several benchmark datasets and show that it outperforms existing self-supervised sequential recommender models. SAMRec allows for improving model generalization to address the data sparsity issue without the self-supervised learning frameworks. 

In summary, our main contributions are as follows:

\begin{itemize}[leftmargin=5mm]
    \item We investigate the Transformer-based recommender models from the perspective of loss geometry and observe that the existing models tend to converge at sharp local minima, resulting in poor training stability and  model generalization.
    \item We propose SAMRec to explicitly smooth the loss geometry during training under the framework of min-max optimization, which improves the efficiency of training and the generalization for the task of sequential recommendation.
    \item The experimental results show our proposed SAMRec outperforms state-of-the-art self-supervised Transformers, such as S$^3$Rec and CL4SRec, without the need for pre-training or strong data augmentations.
\end{itemize}


\section{Related work}
\label{sec:related}
This section briefly reviews the related work on  sequential recommendation and sharpness-aware minimization. 

\subsection{Sequential Recommendation}

Sequential recommendation relies on analyzing sequences of user-item interactions. To infer the probability of an item given the previous items, Markov Chains~\cite{rendle2010factorizing} have been used  to capture the dynamic nature of user behavior. In recent years, the focus has shifted towards utilizing deep neural networks for this task~\cite{HidasiKBT15,tang2018personalized}. Among the various neural networks, Transformer-based models have demonstrated significant potential in sequential recommendation~\cite{kang2018self,lian2020geography,sun2019bert4rec,li2020time,wu2020sse,liu2021augmenting}, due to their ability to model arbitrary dependencies within a sequence.  For example, SASRec~\cite{kang2018self} is a pioneering work adopting the self-attention
mechanism to learn transition patterns in item sequences. Subsequent studies have introduced Transformer variants for different scenarios by incorporating bidirectional attentions~\cite{sun2019bert4rec}, time intervals~\cite{li2020time}, personalization~\cite{wu2020sse}, importance sampling~\cite{lian2020geography}, and sequence augmentation~\cite{liu2021augmenting}. 

However, the effectiveness of Transformer-based models is limited by their dependence on large item corpora for training, which can be problematic when dealing with sparse datasets or short item sequences. To address this issue, researchers have explored the use of self-supervised learning techniques~\cite{zhou2020s3,liu2021augmenting,ma2020disentangled,qiu2022contrastive,xie2022contrastive,chen2022intent}, which typically involve constructing augmentations from unlabelled data and employing a contrastive loss to enhance the discrimination ability of encoders. This loss maximizes agreements between positive pairs, which consists of different augmentations (i.e., views) of the same instance. However, such approaches may encounter various training issues, including initialization sensitivity, additional item features, manual data augmentation (e.g., item crop, item mask, and item reorder), and memory bottlenecks caused by large batch sizes. In this paper, we aim to overcome these limitations by explicitly smoothing the loss geometry of Transformers to improve their model generalization.

\subsection{Sharpness-Aware Minimization}
Many studies have observed that training Stochastic Gradient Descent (SGD) with larger batch sizes can easily lead to test performance degradation. The phenomenon is commonly referred to as the generalization gap \citep{keskar2017on,goyal2017accurate,masters2018revisiting, lin2020extrapolation}, and \citet{keskar2017on} have shown that the degradation of performance is highly correlated with the sharpness of the model parameters. 
This finding has led to many further works which focus on closing the generalization gap \citep{neu2021information, lin2020extrapolation}.
Recently, \citet{Jiang2020Fantastic} build a strong correlation between the sharpness and the generalization error across a wide range of models.
This insight has motivated the idea of minimizing the sharpness during training to improve model generalization, leading to Sharpness-Aware Minimization (SAM)~\citep{foret2021sharpnessaware}. SAM aims to find an optimal solution with low losses across its entire neighborhood, rather than solely focusing on any single point.

The success of SAM has inspired  many subsequent works.
For example, ASAM~\cite{kwon2021asam} introduces adaptive sharpness with a scale-invariant property that adjusts the maximization region of
weight space. FisherSAM ~\cite{kim2022fisher} replaces SAM’s Euclidean balls with
ellipsoids induced by the Fisher information, which obtains more accurate
manifold structures. On the theoretical side, \citet{andriushchenko2022towards} prove that SAM always chooses a solution that enjoys better generalization properties than standard gradient descent for a certain class of problems. \citet{llenhoff2023sam} have established a connection between SAM and Bayes through a Fenchel bi-conjugate. \citet{wen2023how} prove that  the implicit bias of SAM is minimizing the top eigenvalue of Hessian in the full-batch setting. Encouraged by the success of SAM in various learning tasks, we aim to investigate whether SAM can deliver similar gains in sequential recommendation, which does not require pre-training or data augmentations, as in S$^3$Rec \citep{zhou2020s3} and CL4SRec \citep{xie2022contrastive}, to achieve reasonable performance.


\section{Problem and Background}
\label{sec:problem}

\subsection{Problem Formulation}
Let $\mathcal{U}$ be a set of users, $\mathcal{V}$ be a set of items, and $\mathcal{S} = \{s_1, s_2, \dots, s_{|\mathcal{U}|}\}$ be the set of users' actions. Each user $u\in \mathcal{U}$ is associated with a sequence of items in the chronological order represented as $s_{u}= [v_{1}, v_{2}, \dots, v_{|s_{u}|}]$, where $v_{t}\in \mathcal{V}$ denotes the item that user $u$ has interacted with at time $t$, and $|s_{u}|$ is the length of the item sequence $s_{u}$. The goal of sequential recommendation is to predict the next item $v_{|s_{u}|+1}$~\cite{kang2018self,zhou2020s3,xie2022contrastive}.


\subsection{SASRec}

SASRec~\cite{kang2018self} maintains an item embedding table $\mathbf{T} \in \mathbb{R}^{|\mathcal{V}| \times d}$, where $d$ is the dimension of the latent representation for the item. A user sequence $s_u$ can be converted into a fixed-length sequence $(v_1, v_2, \ldots, v_n)$, where $n$ is the predefined maximum length (for example, by truncating or padding items to keep the most recent $n$ items). The embedding for $(v_1, v_2, \ldots, v_n)$ is denoted as $\mathbf{E} \in \mathbb{R}^{n\times d}$, which can be retrieved from the table $\mathbf{T}$. To capture the impact of the position of items in the sequence, a learnable positional embedding $\mathbf{P} \in \mathbb{R}^{n\times d}$ can be injected into the input embedding. The resulting embedding is denoted as $\hat{\mathbf{E}} = \mathbf{E} + \mathbf{P}$, and it can be directly fed to any Transformer-based models:
\begin{equation}
    \mathbf{F} \gets \text{Transformer} (\hat{\mathbf{E}}),
\end{equation}
where Transformer$(\cdot)$ consists of  self-attention layers and point-wise feed-forward layers. Also, one can adopt residual connection, dropout, and layer normalization tricks for stabilizing and accelerating the training (more details can be found in \cite{vaswani2017attention}).

Finally, one can predict the next item (given the first $t$ items) based on $\mathbf{F}_t$, and the  inner product is used to predict the relevance of item $i$ as $r_{i,t} = \langle \mathbf{F}_t, \mathbf{T}_i \rangle$, 
where $\mathbf{T}_i \in \mathbb{R}^d$ is the embedding of item $i$. Recall that the model inputs a sequence  $  (v_1, v_2, \ldots, v_n )$ and its desired output is a shifted version of the same sequence $ (o_1, o_2, \ldots, o_n )$, we  adopt the Binary Cross-Entropy (BCE) loss as:
\begin{equation}
\label{eq7}
\begin{aligned}
\mathcal{L}_{BCE}(\bm{\Theta})= -\sum_{s_u\in \mathcal{S}} \sum_{t=1}^n\left[\log  (\sigma(r_{o_{t}, t}))+  \log (1-\sigma(r_{o'_{t}, t}))\right],
\end{aligned}
\end{equation}
where $\bm{\Theta}$ is the model parameters, $o'_{t} \not\in s_u$ is a negative sample corresponding to $o_{t}$, and $\sigma(\cdot)$ is the sigmoid function. More details can be found in SASRec~\cite{kang2018self}. However, sparsity in the data undermines the capability of a Transformers-based model to learn item correlations in sequences. For example, the model performs poorly on short item sequences~\cite{liu2021augmenting}.

Self-supervised learning has been introduced to alleviate data sparsity issues in Transformer-based sequential models. Next, we introduce two popular approaches: S$^3$Rec~\citep{zhou2020s3} and CL4SRec~\citep{xie2022contrastive}.

\subsection{S$^3$Rec}

S$^3$Rec~\citep{zhou2020s3} improves the architecture by dividing the training into two stages, pre-training and fine-tuning. In the pre-training stage, four self-supervised optimization objectives are utilized to capture item-attribute, sequence-item, sequence-attribute, and sequence-segment correlations:

\begin{itemize}[leftmargin=5mm]
    \item \textbf{Item-Attribute }: It aims to maximize the mutual information between an item and its attributes.
    \item \textbf{Sequence-Item}: It randomly masks a proportion of items,  then recovers them based on the surrounding contexts.
    \item \textbf{Sequence-Attribute}: It recovers the attributes of a masked item based on surrounding contexts.
    \item \textbf{Sequence-Segment }: It further extends the Cloze strategy from a single item to an item segment.
\end{itemize}

During the fine-tuning stage, S$^3$Rec uses the learned parameters from the pre-trained stage to initialize the parameters of the unidirectional Transformer. Then, it adopts the left-to-right supervised signals to train the networks.

\subsection{CL4SRec}

CL4SRec~\citep{xie2022contrastive} incorporates additional self-supervised signals to enhance the user representations through contrastive learning. CL4SRec adopts the following data augmentations:

\begin{itemize}[leftmargin=5mm]
    \item \textbf{Crop}: It randomly selects a continuous sub-sequence of the original sequence.
    \item \textbf{Mask}: It randomly selects a proportion of items to be masked, i.e., replacing them with a special token [mask].
    \item \textbf{ Reorder}: It randomly selects a continuous sub-sequences, then shuffles the items within the sub-sequence.
\end{itemize}

CL4SRec makes use of these three basic augmentations to construct different views of the same sequences and uses contrastive loss to minimize the difference between differently augmented views of the same sequences while maximizing the difference between the augmented views derived from different sequences.

\begin{figure*}
\centering
    \subfigure[SASRec]{\label{fig:Sasrec1}\includegraphics[width=0.18\linewidth]{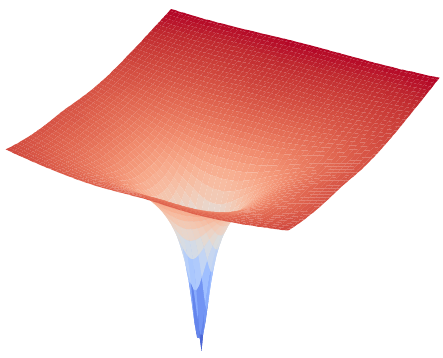}}
    \subfigure[S$^3$Rec]{\label{fig:land-vit}\includegraphics[width=0.18\linewidth]{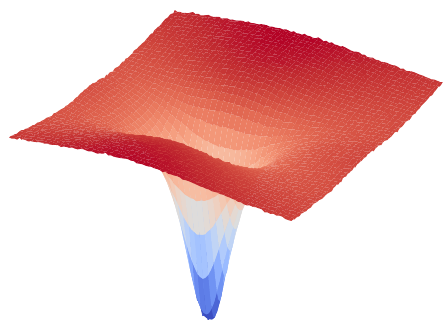}}
    \subfigure[CL4SRec]{\label{fig:land-mixer}\includegraphics[width=0.18\linewidth]{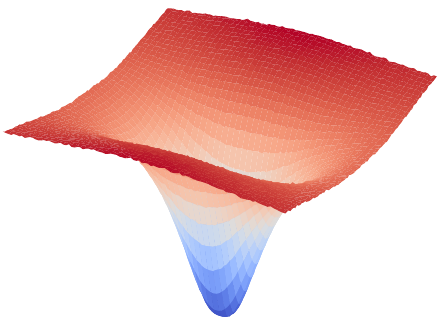}}
    \subfigure[SAMRec]{\label{fig:land-vit-sam}\includegraphics[width=0.18\linewidth]{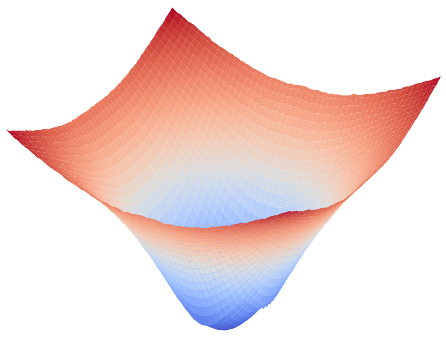}}
\caption{The 3D loss landscapes of SASRec, S$^3$Rec, CL4SRec, and SAMRec.   The sharper the local minima, the less generalizable the model is. SAMRec explicitly smooths the loss geometry during model training by simultaneously minimizing the loss value and smoothing the loss landscape. As such, SAMRec has smoother loss geometry than others.}
\label{fig1}
\end{figure*}

\subsection{ The Sharp Minima Problem}

The connection between the geometry of the loss landscape and generalization has been extensively studied from both theoretical and empirical perspectives \citep{foret2021sharpnessaware, Jiang2020Fantastic, keskar2017on, li2018visualizing}. Thus, we investigate the loss landscapes of SASRec, S$^3$Rec, and CL4SRec to gain insights into their training dynamics. Specifically, given a well-trained model, we calculate the loss values by moving the model parameters $\bm{\Theta}$ along two random directions to generate a 3D loss landscape \citep{li2018visualizing}.

As shown in Figure~\ref{fig1}, we observe that SASRec converges to extremely sharp local minima, whose largest principal curvatures are almost an order of magnitude larger than S$^3$Rec and CL4SRec. Consequently, these sharp minima often lead to poor generalization, since the loss function is very sensitive to model parameters, a small perturbance leads to a dramatic change in the performance. By using self-supervised signals, S$^3$Rec and CL4SRec are able to smooth the loss landscape in an implicit manner. This observation motivates our proposed SAMRec, which explicitly converges to a flat region with small curvature, improving the model performances and model generalization.


\section{The Proposed SAMRec Model}
\label{sec:proposed}

Inspired by Sharpness-Aware Minimization (SAM) \citep{foret2021sharpnessaware,liu2022towards}, we propose SAMRec that explicitly smooths the loss geometry during model training. Formally, SAMRec aims to simultaneously minimize the loss value and smooth the loss landscape, which is achieved by solving the min-max problem:
\begin{equation}
    \min_{\bm{\Theta}} \max_{||\bm{\Delta}||_2\leq\rho} \mathcal{L}_{BCE}(\bm{\Theta}+\bm{\Delta}),
    \label{ee4}
\end{equation}
where $\bm{\Delta}$ denotes the perturbation corresponding to model parameters $\bm{\Theta}$, and $\mathcal{L}_{BCE}(\cdot)$ is the objective of sequential recommendation as given by Eq. (\ref{eq7}). The min-max objective first obtains the perturbation $\bm{\Delta}$ in a neighborhood ball area with a radius $\rho$. The outer optimization tries to minimize the loss of the perturbed weight $\bm{\Theta}+\bm{\Delta}$. Intuitively, our goal is to ensure that small perturbations to the model parameters will not significantly increase the empirical loss, indicating that our SAMRec tends to converge to a flatter minimum, achieving better model generalization.

However, finding the exact optima $\bm{\Delta}^*$ of the inner maximization is challenging, we employ a first-order Taylor expansion as:
\begin{equation}
\label[equation]{eps}
\begin{aligned}
\hat{\bm{\Delta}} & = \mathop{\arg\max}_{||\bm{\Delta}||_2\leq\rho} \mathcal{L}_{BCE}(\bm{\Theta}+\bm{\Delta}) \\
& \approx 
\mathop{\arg\max}_{||\bm{\Delta}||_2\leq\rho} \mathcal{L}_{BCE}(\bm{\Theta}) + \langle \bm{\Delta}, \nabla_{\bm{\Theta}} \mathcal{L}_{BCE}(\bm{\Theta})   \rangle \\
& = \rho \cdot \frac{\nabla_{\bm{\Theta}}\mathcal{L}_{BCE}(\bm{\Theta})}{\|\nabla_{\bm{\Theta}}\mathcal{L}_{BCE}(\bm{\Theta})\|_2}.
\end{aligned}
\end{equation}
We can see that $\hat{\bm{\Delta}}$ is just a scaling of the  gradient with respect to the current model parameters $\nabla_{\bm{\Theta}}\mathcal{L}_{BCE}(\bm{\Theta})$. In this way, the outer minimization is: $\min_{\bm{\Theta}}\mathcal{L}_{BCE}(\bm{\Theta} + \rho \cdot  \nabla_{\bm{\Theta}}\mathcal{L}_{BCE}(\bm{\Theta}) / \|\nabla_{\bm{\Theta}}\mathcal{L}_{BCE}(\bm{\Theta})\|_2)$, which could be implemented by a two-step gradient ascent-descent framework as follows:

\begin{itemize}[leftmargin=5mm]
    \item \textbf{Update} $\bm{\Delta}$: we compute the gradient $\nabla_{\bm{\Theta}}\mathcal{L}_{BCE}(\bm{\Theta})$ at $\bm{\Theta}=\bm{\Theta}^{(t)}$. Then, using Eq. (\ref{eps}) to solve $\bm{\Delta}^{(t)}$ in the maximization, i.e., taking a gradient ascent step of size $\rho$ along the unit vector: $\nabla_{\bm{\Theta}^{(t)}}\mathcal{L}_{BCE}(\bm{\Theta}^{(t)}) / \|\nabla_{\bm{\Theta}^{(t)}}\mathcal{L}_{BCE}(\bm{\Theta}^{(t)})\|_2$.
    \item \textbf{Update} $\bm{\Theta}$: we compute the gradient $\nabla_{\bm{\Theta}}\mathcal{L}_{BCE}(\bm{\Theta})$ at $\bm{\Theta}=\bm{\Theta}^{(t)} + \bm{\Delta}^{(t)}$ for the outer minimization: $\min_{\bm{\Theta}}   {\mathcal{L}_{\text{BCE}}}(\bm{\Theta}+\bm{\Delta})$. The model parameters $\bm{\Theta}$ would be updated via standard gradient descent.
\end{itemize}

We alternately update the perturbations $\bm{\Delta}$ and the model parameters $\bm{\Theta}$  until convergence.\\

\noindent \textbf{Model Complexity}:
Our SAMRec adopts the same model architecture as SASRec with the same level of complexity in terms of parameters. However, the update rule of our SAMRec entails two sequential (non-parallelizable) gradient computations at each step, which could potentially double the computational overhead compared to SASRec. Fortunately, as suggested by~\citet{liu2022towards}, it is sufficient to periodically compute the inner gradient ascent as opposed to calculating it for every single step. This trick reduces the computational complexity of SAMRec significantly while maintaining comparable performance in practice.


\section{Experiments}

In this work, we conduct experiments on four public benchmark datasets: \texttt{Amazon-Beauty}, \texttt{Amazon-Sports}, \texttt{Amazon-Toys}, and \texttt{Yelp}. Following prior studies~\cite{kang2018self,zhou2020s3,xie2022contrastive}, we first keep only the 5-core settings for all datasets and then group the interaction records by users and sort them by interaction timestamps in ascending order. We use the last interacted item of each user sequence for the test set, the second-to-last item for the validation set, and all the earlier items for the training set.

For baselines, we mainly compare our SAMRec with SASRec, S$^3$Rec, and CL4SRec, as our goal is to investigate whether smoothing the loss landscape would improve performance without using any self-supervised learning strategies. We use the default hyperparameters for all the baselines. For SAMRec, we choose SASRec as its backbone and  vary the radius $\rho$ (Eq. (\ref{ee4})) within  $\{0.001, 0.01, 0.1, 1, 10\}$. We use two widely used metrics, Hit Ratio (HR) and Normalized Discounted Cumulative Gain (NDCG), to evaluate performance. We repeat the experiments 10 times independently and report the average results for HR@10 and NDCG@10.

\begin{table*}
\caption{ The performance of different methods in terms of HR@10 and NDCG@10, where he best score are bolded  and the second best are underlined.
Improv denotes the relative improvements our SAMRec over the SASRec.}
\label{t1}
\begin{tabular}{ccccccccc}
\toprule[1.2pt]
         & \multicolumn{2}{c}{Beauty} & \multicolumn{2}{c}{Sports} & \multicolumn{2}{c}{Toys}   & \multicolumn{2}{c}{Yelp}          \\
Model    & HR@10           & NDCG@10         & HR@10           & NDCG@10         & HR@10           & NDCG@10         & HR@10           & NDCG@10         \\ \hline
SASRec   & 0.0563          & 0.0288          & 0.0326          & 0.0194          & 0.0692          & 0.0293          & 0.0333          & 0.0168          \\
S$^3$Rec & 0.0574          & 0.0314          & 0.0335          & 0.0198          & 0.0713          & 0.0304          & 0.0340          & 0.0156          \\
CL4SRec  & \underline {0.0630}    & \textbf{0.0323} & \underline { 0.0351}    & \underline {0.0213}    & \underline {0.0765}    & \underline { 0.0326}    & \underline { 0.0376}    & \underline { 0.0187}    \\
SAMRec   & \textbf{0.0638} & \underline { 0.0320}    & \textbf{0.0362} & \textbf{0.0219} & \textbf{0.0772} & \textbf{0.0330} & \textbf{0.0381} & \textbf{0.0194} \\
Improv.  & +13.32\%        & +11.11\%        & +11.04\%        & 12.89\%         & +11.56\%        & +12.63\%        & +14.41\%        & +15.48\%        \\ \toprule[1.2pt]
\end{tabular}
\end{table*}

\begin{figure*}
	\begin{center}
		\includegraphics[width=15cm]{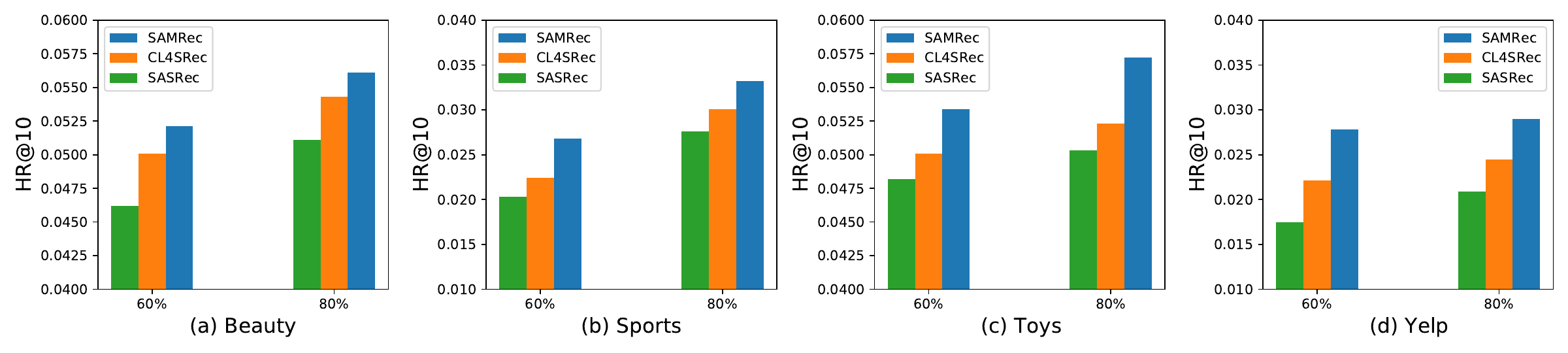}
	\end{center}
	\caption{The results (HR@10) of different models with respect to varying  percentage of training data.}
	\label{hist}
\end{figure*}

\subsubsection*{\textbf{How does SAMRec perform without self-supervised signals?}}

We compare the performance of our SAMRec with all baselines. Table \ref{t1} summarizes the results on the four datasets in terms of HR@10 and NDCG@10. Note that the improvement columns indicate the performance of SAMRec relative to SASRec. From the table, we make two major observations: 1) S$^3$Rec and CL4Rec consistently perform better than SASRec on all datasets, which verifies the effectiveness of self-supervised signals, either from pre-training or data augmentations, in the sequential recommendation task; 2) Our SAMRec generally outperforms all the baselines, except for the Beauty dataset, verifying the benefits of smoothing the loss landscapes of Transformers during training.

Although SAMRec and SASRec have the same Transformer architecture, SAMRec performs better than SASRec, i.e., the improvements of SAMRec over SASRec range from $11.04\%$ to $15.48\%$ in terms of HR@10 and NDCG@10. This is attributed to our training paradigm that prevents the convergence to a sharp local minimum. The first-order optimizers within SASRec (e.g., SGD or Adam) only seek the model parameters that minimize the training error. In contrast, our SAMRec considers higher-order information such as flatness that correlates with generalization performance. On the other hand, our SAMRec has comparable results with self-supervised methods: S$^3$Rec and CL4Rec. In their implementation, both S$^3$Rec and CL4Rec require additional item features, manual data augmentation, and a large batch size for contrastive views, which may encounter training issues. Alternatively, our SAMRec gains from the smoothed loss geometry, without the need for pre-training or strong data augmentations.

\subsubsection*{\textbf{Can SAMRec improve model generalization?}}

Sequential recommendation often encounters data sparsity issues, where users have limited historical records. In order to simulate this scenario, we train a model using only a partial training data set ($60\%$, $80\%$), while keeping the test data unchanged. We compare our proposed SAMRec with SASRec and CL4SRec on all datasets. As depicted in Figure~\ref{hist}, we observe a substantial degradation in performance when less training data is used. However, SAMRec consistently outperforms SASRec and CL4SRec in all datasets. These results confirm our claim that smoothing the loss geometry is able to enhance the data efficiency and generalization of models for the task of sequential recommendation.

\subsubsection*{\textbf{Can smooth loss landscapes help S$^3$Rec and CL4SRec?}}

\begin{table}
\caption{The results of S$^3$Rec and CL4Rec with their variants.}\label{wrap1}
\begin{tabular}{ccccc}
\toprule[1.2pt]
         & Beauty  & Sports  & Toys    & Yelp    \\
Model    & HR@10   & HR@10   & HR@10   & HR@10   \\ \hline
S$^3$Rec & 0.0574  & 0.0335  & 0.0713  & 0.0340  \\
+ SAMRec & 0.0605  & 0.0350  & 0.0756  & 0.0356  \\ 
Improv.  & +5.40\% & +4.78\% & +6.03\% & +4.71\% \\\hline
CL4SRec  & 0.0630  & 0.0351  & 0.0765  & 0.0376  \\  
+ SAMRec & 0.0645  & 0.0365  & 0.0791  & 0.0394  \\
Improv.  & +2.38\% & +3.98\% & +3.40\% & +4.79\% \\\toprule[1.2pt]
\end{tabular}
\end{table}

Our SAMRec can be easily integrated with any Transformer-based models by simply replacing the loss minimization with our min-max optimization as shown in Eq. (\ref{ee4}).  Here we conducted experiments to investigate whether smoothing the loss landscape of S$^3$Rec and CL4Rec would further improve their performance. Table~\ref{wrap1} presents the results in terms of HR@10 for all datasets. From the table, we observe that both variants outperform their original backbones, with improvements ranging from $2.38\%$ to $5.40\%$. These promising results lead us to conclude that explicitly smoothing the loss landscapes contributes to enhancing model performance.

\subsubsection*{\textbf{What is the impact of varying radius $\rho$?}}

\begin{figure*}
	\begin{center}
		\includegraphics[width=15cm]{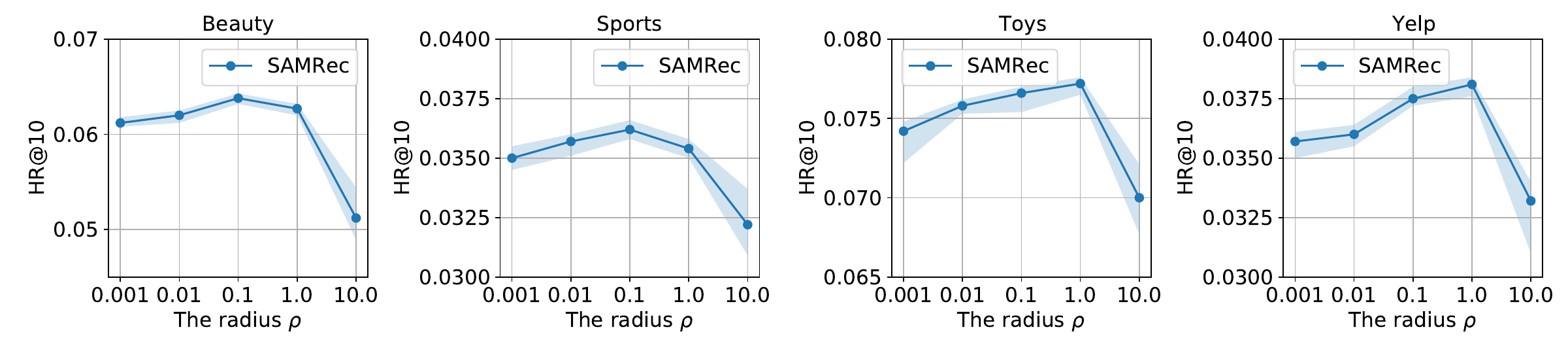}
	\end{center}
	\caption{The impact of $\rho$ for all datasets.}
	\label{aa}
\end{figure*}

Our SAMRec introduces an additional hyperparameter: the radius $\rho$ in Eq. (\ref{ee4}). To analyze the influence of $\rho$, we varied its value in the range of $0.001$ to $10.0$ and reported the experimental results in Figure~\ref{aa}. We observed that our SAMRec remained stable for all datasets when $\rho$ was around the range of $[0.1, 1.0]$. As the radius $\rho$ can be seen as the ascent step for the inner maximization, rather than using a constant value, one can design a scheduler for $\rho$ to make it adaptive to the gradient ascent optimization procedure. We leave this extension as a future direction for exploration.

\section{Conclusion and Future Work}
\label{sec:conclusion}

This paper investigates the loss geometry of several  of Transformer-based sequential recommender models, namely SASRec, S$^3$Rec, and CL4SRec. We observe that  self-supervised learning, employed in S$^3$Rec and CL4SRec, fundamentally altered the loss geometry during model training. As a result, S$^3$Rec and CL4Rec exhibit much smoother and less sharp loss surfaces compared to SASRec. Motivated by this finding, we propose SAMRec, which explicitly smooths out the loss landscape during training. SAMRec achieves superior performance in benchmark evaluations while avoiding the need for data augmentation and pre-training, which are typically required in self-supervised learning approaches. 

For future work, we aim to explore the potential of SAMRec to enhance other models available in the HuggingFace Transformers library. Additionally, we are interested in designing a scheduler for the ascent step (e.g., the radius) to further improve the model performance and reduce the total training time.

\bibliographystyle{ACM-Reference-Format}
\bibliography{reference}

\end{document}